# Drug repositioning for Alzheimer's disease with transfer learning

**Yetao Wu, Han Liu, Jie Yan, Xiaolin Hu**

Department of Computer Science and Technology, Institute for Artificial Intelligence, State Key Laboratory of Intelligent Technology and Systems, BNRist, THBI, Tsinghua University, Beijing, China
xlhu@tsinghua.edu.cn

## Abstract

Deep Learning and DRUG-seq (Digital RNA with perturbation of genes) have attracted attention in drug discovery. However, the public DRUG-seq dataset is too small to be used for directly training a deep learning neural network from scratch. Inspired by the transfer learning technique, we pretrain a drug efficacy prediction neural network model with the Library of Integrated Network-based Cell-Signature (LINCS) L1000 data and then use human neural cell DRUG-seq data to fine-tune it. After training, the model is used for virtual screening to find potential drugs for Alzheimer's disease (AD) treatment. Finally, we find 27 potential drugs for AD treatment including Irsogladine (PDE4 inhibitor), Tasquinimod (HDAC4 selective inhibitor), Suprofen (dual COX-1/COX-2 inhibitor) et al.

## 1 Introduction

Alzheimer's disease is a common, complex, neurodegenerative disease(Lee and Kim 2020), which is extremely challenging for drug development. Drug repositioning is attractive to AD drug development because toxicity, pharmacokinetics, and pharmacodynamics profiles of a given drug are fully characterized(Kwon et al. 2019). Recently, deep learning methods have shown great potential for drug discovery(Jang and Cho 2019) and repositioning(Pham et al. 2022; Pham et al. 2021; Zhu et al. 2021). In particular, several research groups have predicted drug efficacy from the L1000 dataset(Subramanian et al. 2017). It contains gene expression profiles which includes the responses of different compound treatments (Pham et al. 2022; Pham et al. 2021; Zhu et al. 2021).

Despite being widely used, there are three common problems encountered when utilizing the L1000 dataset. First, most of the L1000 data is obtained from tumor cell lines which carry many mutations. When the L1000 data is applied to predict the gene expression profile of normal cells, the context may be different. Second, as the L1000 method and commercial services based on it are not prevalent, it is difficult to get customized L1000 data for specific cell type. Third, the L1000 dataset contains many unreliable and noisy gene expression profiles (Pham et al. 2021).

Nowadays RNA-seq methods and commercial services are very prevalent. Based on RNA-seq methods, DRUG-seq is a reliable and cost-effective tool for comprehensive transcriptome readout in high-throughput drug screening, which costs 2 – 4 US dollar per sample(Ye et al. 2018). However, since DRUG-seq is a new technique, there are no large public DRUG-seq datasets yet. It is very difficult to train a deep learning neural network from scratch with a small public DRUG-seq dataset. Inspired by the transfer learning technique (apply knowledge gained in one task to a related task), we propose to utilize transfer learning for reducing the amount of training data needed when training neural networks with the DRUG-seq dataset.

Since the DRUG-seq dataset and the L1000 dataset both consist of gene expression profiles (Jeong et al. 2017) which cover the responses of different compound treatments, we can take advantage of the similarity and correlation between the DRUG-seq dataset and the L1000 dataset for transfer learning. In this study, we pretrained a drug efficacy prediction neural network model with the L1000 data and then used human neural cell DRUG-seq data to fine-tune it. After training, the model was used to predict profiles for new chemicals in the DrugBank database. These profiles were then used for virtual screening to find potential drugs for Alzheimer's disease treatment. To the best of our knowledge, this is the first time that the DRUG-seq data has been used in deep learning.

## 2 Methods

**Datasets**

In the following paragraphs we introduce several datasets used in this study, including STRING, DrugBank, L1000, DRUG-seq dataset of human neural cells, and transcriptome data of Alzheimer's disease patients.

*The High-quality L1000 dataset*

According to the previous literature(Pham et al. 2021; Qiu et al. 2020), experiments were conducted on a selected high-quality L1000 dataset in this study. Because the original L1000 dataset contains many unreliable and noisy gene expression profiles, a much better prediction performance was obtained by using the selected high-quality L1000 dataset than by using the original L1000 dataset (Pham et al. 2021). The selected high-quality L1000 dataset consists of 1944 training samples (284 chemicals), 556 developing samples (92 chemicals) and 502 testing samples (92 chemicals) (Pham et al. 2021).

*The STRING database*

STRING is a database of known and predicted protein–protein interactions, including direct and indirect associations(Pham et al. 2021; Szklarczyk et al. 2019). The human protein–protein interaction network, which consists of approximately 12,000,000 interactions (edges) and 19,000 proteins (nodes), was extracted from the STRING database to compute vector representations for 978 L1000 genes in a previous study(Pham et al. 2021). And the drug-target vector representations used in this study were also computed from the STRING database(Pham et al. 2021; Szklarczyk et al. 2019). The details of generating these representations were presented in previous literature (Pham et al. 2021).

*The DRUG-seq dataset*

We used the DRUG-seq data of a previous study(Rodriguez et al. 2021), in which human neural cells were treated with compounds or DMSO for 24h. The number of samples used was 661, 75% of which was used for training. Raw sequencing data can be found at the National Center for Biotechnology Information (NCBI) Sequence Read Archive (SRP301436)(Rodriguez et al. 2021). Processed data can be found on Gene Expression Omnibus (GSE164788) (Rodriguez et al. 2021).

*Expression profiles (RNA-Seq)*

Expression profiles from both Alzheimer's disease patients and healthy negative controls were obtained from two previous studies(Mizuno et al. 2021; Nativio et al. 2020).

The expression profiles data which we used firstly was downloaded from the NCBI (GSE173955)(Mizuno et al. 2021). Post mortal human hippocampus from 8 AD and 10 healthy subjects were processed with Illumina TruSeq stranded mRNA LT Sample Prep kit, and then the sequences were obtained with HiSeq1500 according to the manufacturer's protocol(Mizuno et al. 2021). Details of differential expression of gene level between AD and non-AD hippocampi can be found in a previous study(Mizuno et al. 2021). Not all 978 L1000 genes appeared in the result. And therefore, only genes that appeared in both the L1000 dataset and the differential expression gene list were considered when comparing with drug-induced gene expression profiles.

The expression profiles data which we used secondly was downloaded from the NCBI(GSE159699)(Nativio et al. 2020). Post mortal human hippocampus from 10 AD and 12 non-AD control old subjects were processed with the NEBNext Ultra Directional RNA library Prep Kit for Illumina (NEB), and then the sequences were obtained with NextSeq 500 Platform (Illumina) according to the manufacturer's protocol(Nativio et al. 2020).

*The DrugBank database*

The DrugBank database which consists of information about 11,179 drugs and their targets (Pham et al. 2021; Wishart et al. 2006) is a famous, comprehensive, freely accessible database used in many cheminformatics and bioinformatics tasks(Wishart et al. 2006). In this study, we predicted gene expression profiles for drugs in DrugBank. And then they were used to screen potential drugs for Alzheimer's disease treatment.

**DeepCE**

DeepCE is a neural network-based model implemented with python for the 978 L1000 gene expression prediction tasks(Pham et al. 2021), which utilizes multihead attention mechanism and graph neural network. The inputs of the DeepCE model are information about chemical, dosage, cell-type, and the output of this model is gene expression profile. The simplified molecular-input line-entry system (SMILES) strings are used for describing the structure of chemical. Six kinds of dosages inputs are from 0.04 μmol/L to 10 μmol/L. Seven kinds of cell types inputs are allowed in the original DeepCE model: 'A375', 'HA1E', 'HELA', 'HT29', 'MCF7', 'PC3', 'YAPC'(Pham et al. 2021). They were handled by one-hot encoding. Because we wanted to fine-tune this model with DRUG-seq dataset for human normal neural cells, we set 'neural cells' as the eighth cell type for the DeepCE model before training.

Details of the DeepCE model can be found in previous literature (Pham et al. 2021). Its source code and usage instructions can be found in Zenodo (https://doi.org/10.5281/zenodo.3978774) and GitHub (https://github.com/pth1993/DeepCE).

**Transfer learning pipeline**

First, we pretrained the DeepCE model with the high quality version of L1000 dataset in a supervised manner (100 epochs). Second, we fine-tuned this model with DRUG-seq dataset of human neural cells in a supervised manner (25 epochs) (Figure 1). The Xavier initialization(Glorot and Bengio 2010) and the Adam optimizer(Kingma and Ba 2014) were used. The learning rate was 0.0002. The batch size was 16. The dropout rate was set to 10%.

**Baseline model**

The DeepCE model trained from scratch with DRUG-seq dataset of human neural cells was used as the baseline model (25 epochs).

**Performance evaluation**

The Spearman's rank-order correlation score which measures the relationship between predicted gene expression profiles and ground truth was used as the main metric to evaluate performances of models in our study (Pham et al. 2021).

## 2 Results

DRUG-seq is a new technique, and there are no large public DRUG-seq datasets yet. For instance, only 661 samples in the DRUG-seq dataset of human neural cells (Rodriguez et al. 2021) are available. However, training deep learning neural networks from scratch requires a lot of data. Because the DRUG-seq dataset of human neural cells is too small, we could not train a model to predict relative gene expression profiles which caused by perturbation. So, we trained this model with gene expression profiles which were much easier for prediction. Predicted gene expression profiles are then divided by the DMSO control group gene expression profiles to get the relative gene expression profiles.

First, we tried to directly train the DeepCE model with the DRUG-seq dataset of human neural cells (baseline model, 25 epochs), the test set Spearman correlation coefficient was 0.2556. Therefore, we may need to use the transfer learning method to improve the performance of the DeepCE model. First, we pretrained the DeepCE model with the high-quality version of L1000 dataset (100 epochs)(Pham et al. 2021; Subramanian et al. 2017). And then, we fine-tuned this model with the DRUG-seq dataset of human neural cells (25 epochs)(Rodriguez et al. 2021) (Figure 1). This transfer learning pipeline obtained a Spearman correlation coefficient of 0.4301 on the test set, which outperformed the baseline model (25 epochs). Because the DRUG-seq dataset hasn't been reported to be utilized in deep learning before, there is no state-of-the-art (SOTA) method for the DRUG-seq dataset to be compared with our method. Due to the similarity in the context, our model might be more useful for drug repositioning for Alzheimer's disease than the SOTA method for the L1000 dataset which did not use the DRUG-seq dataset for human normal neural cells.

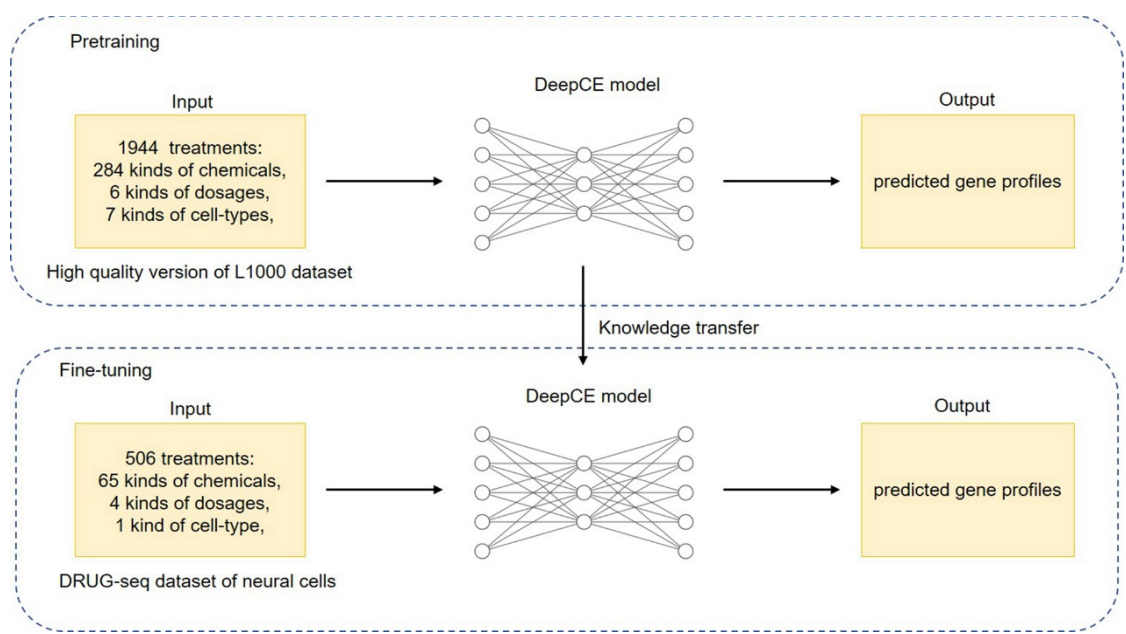

**Fig.1** Transfer learning pipeline for the DeepCE model. The DeepCE model was pretrained with the high-quality version of L1000 dataset (1944 training samples) and then fine-tuned this with DRUG-seq dataset of human neural cells (506 training samples). The inputs of the DeepCE model are information about chemicals, dosages, cell-types, and the output of this model is predicted gene expression profiles (978 dimensions).

We used the trained DeepCE model to do a virtual screen for potential AD drugs according to previous study(Pham et al. 2021). The drug dosage was set to 10μmol/L and the cell-type was set to "neural cells". Differential expression of gene level between AD and non-AD hippocampi can be found in previous literature (Mizuno et al. 2021). We then screened the DrugBank database by computing Spearman's rank-order correlation scores between their gene expression signatures with the AD patient gene expression signatures, and selected the top 10 drugs with the most negative scores as the potential drugs for AD treatment (Table 1) (Mizuno et al. 2021; Pham et al. 2021).

**Table 1 | The chemical structures, status and known uses of potential drugs for AD treatment in the first screen (order by rank)**

| Drug | Structure | Status | Known uses |
|---|---|---|---|
| Irsogladine | 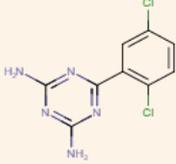 | Investigational | PDE4 inhibitor |
| Thiosalicylic acid | 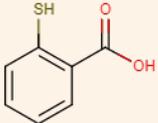 | Experimental | Not Available |
| Homoserine Lactone | 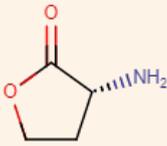 | Experimental | Target Sodium/hydrogen exchanger 1 |
| AZD-7325 | 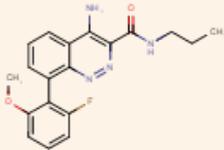 | Investigational | Modulator of the GABAA receptor |
| Tasquinimod | 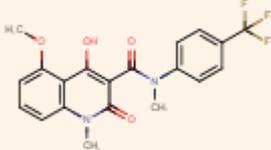 | Investigational | Anti-angiogenic |
| AZD-2066 | 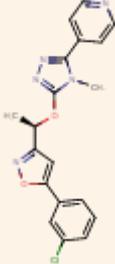 | Investigational | Pain |
| (3-Carboxy-2-(R)-Hydroxy-Propyl)-Trimethyl-Ammonium | 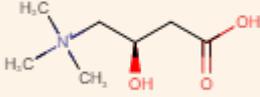 | Experimental | Hyperlipoproteinemias |
| Leucine | 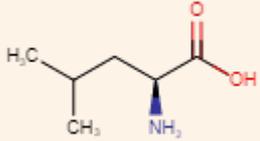 | Investigational, Nutraceutical | Nutrition |
| Diphenidol | 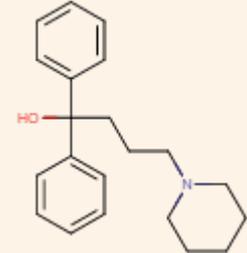 | Approved, Investigational, Withdrawn | Antiemetic agent |
| 25-Hydroxycholesterol | 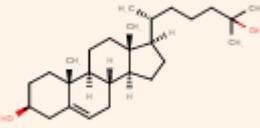 | Experimental | Not Available |

Some of targets of the candidates were reported to be closely related with AD. For instance, the rank 1 potential drug for AD treatment in the screen is Irsogladine. It is a PDE4 inhibitor(Kyoi et al. 2004), and PDE4 inhibitors have been shown to be effective in enhancing the memory and cognition in AD(Tibbo et al. 2019). The rank 5 potential drug for AD treatment in the screen is Tasquinimod. It is an HDAC4 selective inhibitor. Treatment with Tasquinimod significantly upregulated the expression of neuronal memory related genes(Chen et al. 2021).

Since differential expression of gene level between AD and non-AD hippocampi is different in previous literature, we also did a virtual screen for potential AD drugs with expression profiles data from another previous literature(Nativio et al. 2020) (Table 2). The rank 2 potential AD drug in the screen is Erteberel. It is an estrogen receptor beta agonist, which was reported to improve the spatial recognition memory in previous study(Zhao et al. 2013; Zhao et al. 2015).

**Table 2 | The chemical structures, status and known uses of potential drugs for AD treatment in the second screen (order by rank)**

| Drug | Structure | Status | Known uses |
|---|---|---|---|
| 6-(N-Phenylcarbamyl)-2-Naphthalenecarboxamidine | 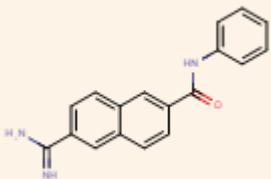 | Experimental | Not Available |
| Erteberel | 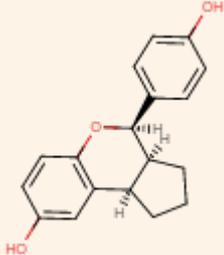 | Experimental, Investigational | Estrogen receptor beta agonist |
| Quarfloxin | 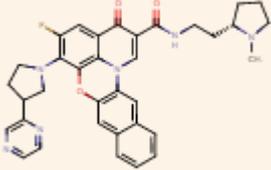 | Investigational | Cancer |
| Chlorophetanol | 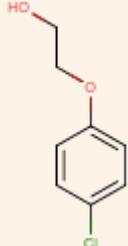 | Experimental | Antifungal |
| 2-Oxy-4-Hydroxy-5-(2-Hydrazinopyridine) Phenylalanine | 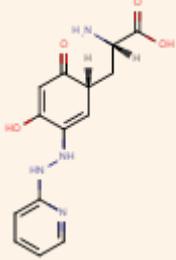 | Experimental | Not Available |

| Oxogluric acid 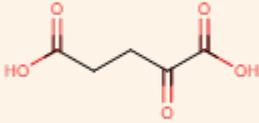 | Experimental, Investigational, Nutraceutical | Intermediate of the Krebs cycle |
| --- | --- | --- |
| 1-naphthaleneacetic acid 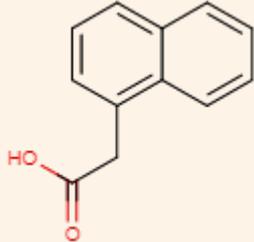 | Experimental | Digestive problems |
| Benzylfentanyl 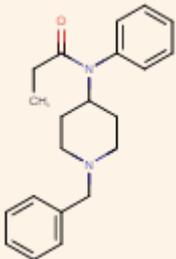 | Illicit | Bind the mu opioid receptor |
| Solamargine 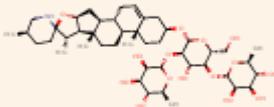 | Investigational | Actinic Keratosis |
| RU84687 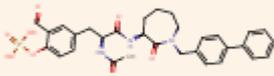 | Experimental | Src SH2 selective binder |

Then we took the intersection, and found several drugs appear in top 500 candidates both in the two virtual screens (Table 3). One of them is Amitriptyline, which is an FDA-approved tricyclic antidepressant. It can significantly benefit AD brains(Chadwick et al. 2011). Potential functions of Antidepressants in the treatment of AD were reported in previous literature(Kim et al. 2013). Another candidate is Suprofen, which is a dual COX-1/COX-2 inhibitor. It was reported that Cyclooxygenase inhibition could prevent memory impairment in Alzheimer's disease model mice(Woodling et al. 2016).

Table 3 | potential drugs appear in top 500 candidates both in the two virtual screens

| Drug | Structure | Status | Known uses |
|---|---|---|---|
| Homoserine Lactone | 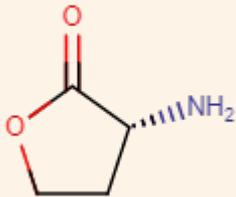 | Experimental | Target Sodium/hydrogen exchanger 1 |
| (2Z)-N-biphenyl-4-yl-2-cyano-3-hydroxybut-2-enamide | 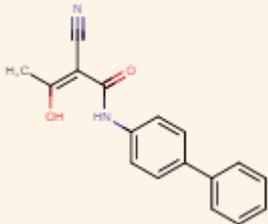 | Experimental | Target Dihydroorotate dehydrogenase (quinone) |
| Amitriptyline | 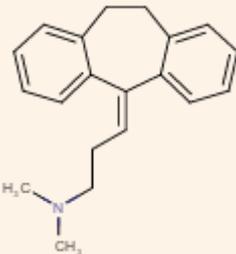 | Approved | Depressive illness |
| Phosphate ion | 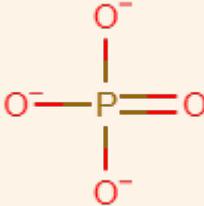 | Experimental | Not Available |
| Rifapentine | 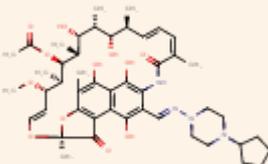 | Approved, Investigational | Antibiotic |
| 4-(4-Hydroxy-3-Isopropylphenylthio)-2-Isopropylphenol | 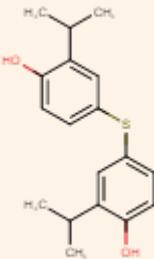 | Experimental | Not Available |
| Suprofen | 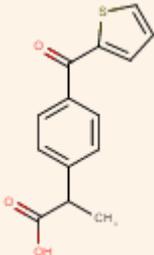 | Approved, Withdrawn | Ibuprofen-type anti-inflammatory, Analgesic, Antipyretic |

## 4 Discussion

Deep Learning and DRUG-seq have attracted attention in drug discovery (Pham et al. 2022; Ye et al. 2018). However, it is very challenging to directly train a deep learning neural network with a small public DRUG-seq dataset. In this study, we pretrained a drug efficacy prediction neural network model with the L1000 data and used the human neural cell DRUG-seq data to fine-tuned it. After training, the model was used for virtual screening to find potential drugs for AD treatment. To the best of our knowledge, this is the first time that DRUG-seq data is used in deep learning.

In the virtual screening, several targets (PDE4, HDAC4, COX-1/COX-2) of the candidates which were reported to be closely related with AD might indicate that our method is useful. More of the candidates were not reported to be related with AD treatment before, which might be new clues for AD research.

Since DRUG-seq is a new method, the available public DRUG-seq dataset is too small. We could not train a model to predict relative gene expression profiles which caused by perturbation. So, we trained this model with gene expression profiles which were much easier for prediction. Predicted gene expression profiles are then divided by the DMSO control group gene expression profiles to get the relative gene expression profiles. This study can only be regarded as proof-of-concept research. In the future, if larger public DRUG-seq datasets become available, we could train a model to predict relative gene expression profiles and get a better result with this transfer learning pipeline.

## 5 Conflict of interest

The authors declare that they have no conflict of interest.

## 6 Acknowledgements

This work was supported by the National Natural Science Foundation of China (Nos. 62061136001, U19B2034, 61836014).